# Photonic machine learning implementation for signal recovery in optical communications

Apostolos Argyris*,[1], Julián Bueno[1], Ingo Fischer[1]

[1] Instituto de Física Interdisciplinar y Sistemas Complejos IFISC (CSIC-UIB), Campus UIB, 07122, Palma de Mallorca, Spain

* email: apostolos@ifisc.uib-csic.es

**Machine learning techniques have proven very efficient in assorted classification tasks. Nevertheless, processing time-dependent high-speed signals can turn into an extremely challenging task, especially when these signals have been nonlinearly distorted. Recently, analogue hardware concepts using nonlinear transient responses have been gaining significant interest for fast information processing. Here, we introduce a simplified photonic reservoir computing scheme for data classification of severely distorted optical communication signals after extended fibre transmission. To this end, we convert the direct bit detection process into a pattern recognition problem. Using an experimental implementation of our photonic reservoir computer, we demonstrate an improvement in bit-error-rate by two orders of magnitude, compared to directly classifying the transmitted signal. This improvement corresponds to an extension of the communication range by over 75%. While we do not yet reach full real-time post-processing at telecom rates, we discuss how future designs might close the gap.**

Recent developments in neuro-inspired information processing using recurrent neural networks (RNNs), cognitive computing approaches, machine learning techniques and deep learning[1,2] architectures have had a major impact on solving classification and pattern recognition tasks with remarkable efficiency[3,4,5,6,7]. However, there are hardly any solutions available if the task is time-dependent, the speed requirements are very demanding and the signals to be processed are of high complexity. To this end, analogue hardware implementations of these information processing tools have been gaining increasing recognition[8]. In recent years, implementations of feed-forward and recurrent NNs based on extreme learning machines (ELM)[9,10] and reservoir computing (RC) approaches[11,12,13,14] have been presented in optoelectronic[15,16,17,18] and photonic[19,20,21,22,23,24,25] hardware. These implementations were in some cases assisted by field programmable gate array (FPGA) modules[25,26]. So far, they have only been employed for standard benchmark tasks such as pattern classification, speech recognition, nonlinear time series prediction and wireless channel equalization. Evolving these hardware implementations to minimal conceptual complexity and to maximal speeds would enable to address signal processing tasks in critical technological fields. An excellent example with ultra-fast post-processing requirements can be found in the contemporary fibre-optic communication networks that now operate even beyond the Tb/s scale[27]. The technological advances in this field target on the highest data throughput over the longest distances with energy efficient and low complexity designs. However, transmission impairments[28], such as chromatic dispersion, Kerr effect and four-wave mixing, put strict limitations on communication speed and distance in fibre-optic communication systems. Current research aims at extending these limits, by focusing mainly on the two ends of the communication links. At the transmitter side, major efforts target on optimizing the emitter[29,30], as well as the encoding communication scheme, by using multi-level formats and signal shaping[28,31,32,33,34]. At the receiver side, high-speed digital signal processing (DSP) algorithms[35,36,37,38,39] with low-complexity have improved signal recovery by mitigating linear and nonlinear signal distortions. The aforementioned approaches in fiber-based communication systems currently shape the status quo of the field, but they are also facing challenges for future trends. For example, the current DSP methods are efficient as long as nonlinear signal distortions do not become too complicated. For this reason, optimal designs of various transmission systems dictate that the launched optical power in standard single mode fibres (SSMF) should be always restricted to moderate levels (around or below 1mW). Inevitably though, these power levels limit the signal-to-noise ratio (SNR) of the received signal, given the standard detection capabilities of fast photoreceivers. In a reasonable consideration, one could suggest to increase the launched optical power into the fibre. There are numerous semiconductor laser emitters available ready to offer tens of mW of emission at telecom wavelengths. Such signals exhibit higher optical SNR that could lead to increased transmission distance, but at expense of enhancing the nonlinear behaviour of the transmission line. Travelling signals will undergo a more complex nonlinear transformation, and eventually it will be too difficult to identify and interpret them at the receiver. Only lately, machine learning algorithms have been in the spotlight of the optical communications community[40]. They are being considered for optical network monitoring and optimization[41,42,43], optical header recognition[44] and mitigation of transmission effects[45,46,47,48,49,50,51,52,53,54]. Still, the drawback of applying these standard tools in ultrafast systems is that they are computationally expensive and still far away from reaching real-time processing at telecom data rates.

In the present work, we provide a first validation that neuro-inspired information processing based on photonic implementations can address critical issues in the field of signal processing for high-speed communications. Specifically, we demonstrate that techniques like ELM and RC can offer solutions to data recovery of distorted signals from extended fibre transmission. We introduce a simplified RC approach with a sequential data processing architecture that allows for a high-speed hardware implementation. Our focus in this work is to efficiently classify signals that have undergone a significant nonlinear transformation with time dependencies. The concept we demonstrate here is generic and powerful and it can be applied to signals that may originate from any optical communication system configurations (different information modulation formats and communication speeds). The obtained results of RC-based processing yields very promising performance, even if it does not yet reach the status of well-established methodologies in signal processing, such as maximized-likelihood sequence detection and back-propagation DSP. However, it represents a revolutionary tool for fast signal processing of problems with increased complexity. RC implementations have already achieved comparable performance to digital algorithms, which are based on Volterra series filters, applied in equalization tasks for nonlinear satellite communication channels[55], however at much slower speeds. Future works that will compare cognitive computing approaches with current DSP technology will be also of significant interest.

**A simplified reservoir computing concept**

The originally proposed RC concept[11,12] for data processing consists of three layers: the input, the reservoir and the output layer. The reservoir is described as a recurrent network with randomly connected nonlinear nodes. Its role is to nonlinearly transform the input and, at the same time, to generate a mapping of the input onto a high-dimensional state space. Recently it was shown[13] that a single nonlinear element with time-delayed feedback can emulate a recurrent network by defining multiple nodes within the feedback loop with delay time $\tau$. These nodes –also denoted as virtual nodes – form a unidirectional ring topology. The output of the single nonlinear element, sampled at time intervals separated by $\theta$, results in a vector of the virtual node values, which is interpreted as the state of the virtual network. When $\theta<T$, with $T$ being the characteristic time of the nonlinear element, the state of each virtual node has, due to inertia, cross-talk with the state of its neighbours, increasing the connectivity among the nodes. Due to the delayed feedback topology, the network states are additionally influenced by their past states, one time-delay before. The final network size $N$ is defined as the number of virtual nodes along the reservoir loop ($N=\tau/\theta$). The structure of the input layer defines how the initial information is fed into the reservoir. For temporal information processing, the simplest way to configure the input interface is to have time-multiplexed inputs. For scalar data inputs, one commonly injects sequentially each input value into the $N$ virtual nodes, within one time interval $\tau$. In order to obtain a large number of different transient responses for this input value, temporal masking is applied. The masking sequence is a length $N$ vector of usually random values and is repeated for every interval $\tau$. This mask defines the input weights for each virtual node. In this way, one masked input value is represented by a $1xN$ vector within one interval $\tau$. This procedure generates a representation of the input in a state space of increased dimensionality before injecting it into the reservoir. When going from scalar data inputs to data vectors with multiple components, the mask is usually replaced by a random connectivity matrix. This matrix defines how the vector components of the input data are fed into the reservoir's nodes.

The information processing technique we propose is shown in Fig. 1, illustrating the simplified RC concept we will explain below. In our implementation, the input information is a one-dimensional data vector, constituted by a number of samples within one bit duration. The motivation is that in all physical communication systems, the digital information of each bit $b^i$ includes a pattern that can be described by a vector $a^i = \{a_1^i\ a_2^i\ a_3^i\ ...\ a_j^i\}$ with $j$ samples within the bit period, as shown in Fig. 1a. The actual sampling rate of the communication system defines the dimension of the vector $a^i$. Thus, $j$ input values represent one bit of digital information per time interval $\tau$. If we followed the original approach, we would need to inject these data vectors into the delay reservoir after multiplying them with the random connectivity matrix. This represents a significant complication when implementing the scheme in hardware and at high speeds. Therefore, we introduce a conceptual simplification which still yields excellent results. We choose $j$ to be equal to the number of virtual nodes $k$. Then we take the elements of the input vector $a_j^i$ in the order defined by the input data stream and multiply them by the random mask $m = \{m_1, m_2, m_3, ...\ m_k\}$. This is illustrated in Fig. 1b. The dimension of the space spanned by the components of the input samples is the same as the dimension of the space defined by the nonlinear responses of the reservoir within one interval $\tau$. As we will show, the concept works very well since the dimensionality of the representation of the data in these two spaces can differ significantly. We note that the flexibility of our approach allows also that the value $k$ can be chosen differently from the number of samples $j$; $k$ can even be different from the total number of nodes $N$ which are available within one reservoir's time delay $\tau$. The importance of selecting these values will be discussed in detail in the results section.

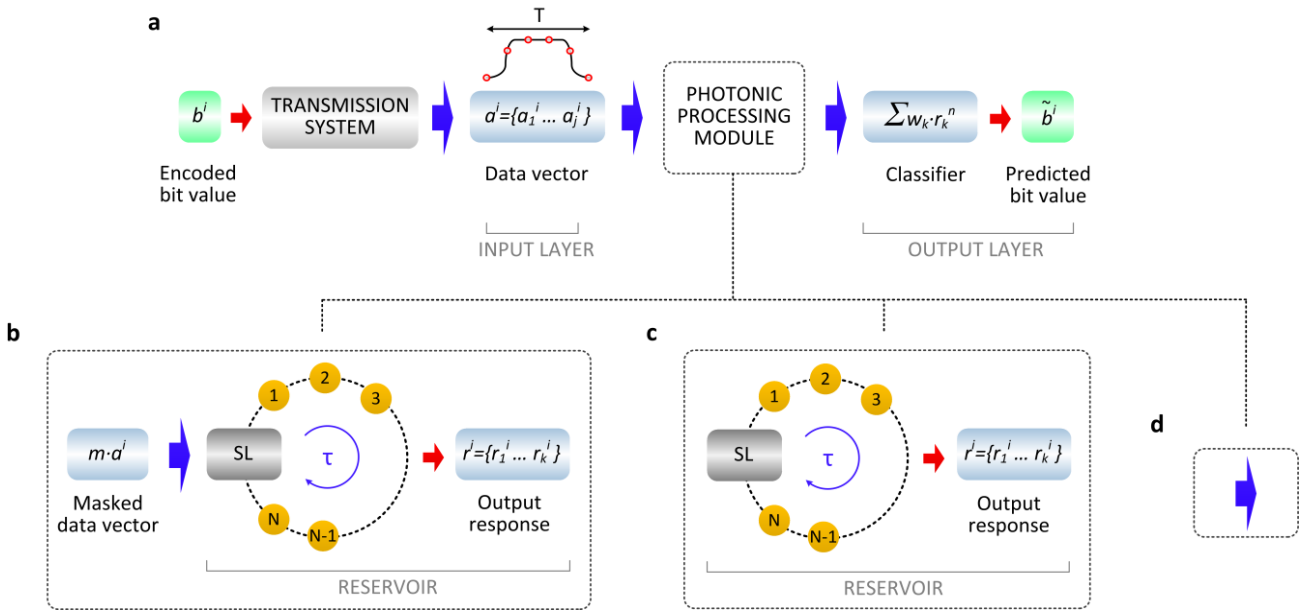

**Figure 1 | Architecture of the photonic machine learning systems for recovering distorted optical communication signals. (a)** A stream of random bits $b^i$ – each of which is represented by an analogue pattern $a_j^i$ with $j$ samples – are transmitted in a fibre-optic transmission system. The detected patterns are used to train a linear classifier. **(b)** An intermediate processing step is deployed in the hardware domain, by introducing a photonic reservoir that consists of a SL subjected to time-delayed feedback. The delay loop of the reservoir is divided into $N$ intervals that define the virtual nodes of a neural network. Each input pattern is multiplied by the same random mask vector $m$, before it is injected into the reservoir (Methods). The pattern $r$ that emerges from $k \leq N$ transient responses of the reservoir nodes is used to train the classifier (Methods). If the classifier takes into account also the responses of neighbouring bit timeframes, the number of transient responses used for classification will be $n \cdot k$, where $n$ is the total number of considered bit

timeframes. We benchmark the reservoir's performance by comparing it to the same linear classifier that is trained on **(c)** the reservoir output without masking or **(d)** the direct output signal from the transmission system.

The output layer performs the readout of the reservoir and is implemented in the originally proposed way (Fig. 1a). The responses of the virtual nodes are linearly combined using a set of optimal readout weights (Methods). The latter are determined by a training process (ridge regression algorithm) that minimizes the mean square error between the obtained and the desired output of an initially defined training sequence. After defining the optimal readout weights, these are kept unaltered during the complete information processing task. This means that the training process is applied once in order to define the readout weights. Finally, a linear summation of the weighted reservoir's responses generates a single computational result per time delay[12,20] which is the prediction output. In the following, the output layer is trained to provide an estimation on the initial binary input ($\tilde{b}^i$).

For the implementation of the reservoir we consider a photonic topology that consists of a semiconductor laser (SL) with time-delayed optical feedback of controllable strength[20] and time-dependent external optical injection that carries the input information (Fig. 1b). In such a system, the identification of the characteristic time *T* of our nonlinear element is a nontrivial task. The definition of a single characteristic time for the system based on the solitary SL alone would be misleading and inaccurate. Depending on conditions related to laser biasing, optical feedback strength, feedback time-delay, optical injection strength and time-scales of the injected external signal, the characteristic time of the response system can be well above the relaxation oscillation bandwidth of the solitary laser. A solitary SL exhibits inherently a dynamical operating bandwidth of several GHz, which is particularly useful for generating fast transient states when injecting external perturbations. This value can further increase up to tens of GHz, as it has been shown in various schemes that SLs under optical injection can exhibit bandwidth enhanced responses[56,57]. In our experiment, depending on the operating conditions of the reservoir and the given SNR that we have available for signal detection, we observe characteristic frequencies $T^{-1}$ of the reservoir response of up to ~ 10GHz. Accordingly the virtual node spacing *θ* was selected to be 50ps.

A similar topology is also used to implement the ELM approach, which is obtained by simply minimizing the optical feedback strength in the reservoir's loop[10]. In our investigations, the performance of the RC and ELM topologies is benchmarked against simplified (no masking of the input, Fig. 1c) or absent (Fig. 1d) photonic processing.

**Input signal considerations**

The versatile strategies that have been developed for the different needs of communicating counterparts – mainly distinguished by the coverage range – have promoted a diversity of systems. We will demonstrate that our approach is generic and applicable on signals with deterministic distortion that emerge from differently structured transmission systems. We focus here our investigations on two fundamental systems with different ranging applications that operate at the telecom C-band (1550nm) and use a single channel carrier (Methods). The first is a short-reach transmission system in which the use of any inline component (dispersion compensation fiber - DCF, optical amplification) is avoided. This type of connection is advantageous for data centre intra- and inter- connections[58], as well as for the next generation DCF-free metro networks, minimizing the cost and complexity of the communication[59]. For this system we consider a non-return-to-zero (NRZ), pulse amplitude modulation (PAM) encoding at a bit rate of $R_1$=25Gb/s (Supplementary Fig. 1), a target bit-rate for the IEEE 802.3 communication standards[60]. The task of the RC process for this system is to mitigate two coexisting phenomena: a linear distortion caused by chromatic dispersion and a nonlinear distortion caused by the Kerr effect. The second system we investigate is a long-haul transmission link with the same encoding, at $R_2$=10Gb/s, including dispersion post-compensation and filtered optical amplification every 100km span (Supplementary Fig. 2). The task of the RC process for this system is to mitigate the Kerr nonlinearity in presence of stochastic noise that originates from the optical amplification modules. In both systems we consider high-power launched optical signals (10mW), which are usually not favoured in conventional transmission systems. In this way we obtain a high optical SNR (36.6dB, more than 10dB higher than conventional systems) for the transmitted signals that allows us to extend the SSMF transmission distance $z$. Even if the data recovery bit-error-rate (BER) becomes higher than 0.1, by considering incoherent detection in absence of any DSP, this is not due to the limited optical SNR but due to deterministic effects that can be in principle compensated for. Specifically, in the investigated scenario for the short-reach transmission we consider a transmission length of $z_1$=45km (Supplementary Fig. 3, Fig. 4). In the long-haul transmission scenario, we consider a transmission length of $z_2$=4000km (Supplementary Fig. 3, Fig. 4). Both systems have been numerically simulated using the nonlinear Schrödinger equation (NLSE) model[28] (Methods), adjusted to include also effects such as polarization mode dispersion and inter-channel nonlinear effects (four wave mixing, cross-phase modulation). The latter do not apply in our investigation with a single transmission channel. The output signals after simulating transmission and photodetection are used to feed the input of the experimentally-built photonic reservoir. Other modulation formats (eg. PAM4) and at higher data bit rates (eg. 56Gb/s) have been also tested, with analogous performance, but their description goes beyond the scope of this manuscript.

**Implementation of the photonic reservoir**

The implemented photonic reservoir is presented in Fig. 2a, following the simplest possible experimental topology. Details for the experimental implementation can be found in Methods. The reservoir consists of a 1542nm discrete-mode quantum-well SL (response laser), and an optical fibre delay loop ($\tau$=66.000 ± 0.025ns) from which the laser receives delayed feedback. A distributed feedback (DFB) SL (injection laser) is used to generate the optical carrier that carries the masked signal from the simulated transmission systems into the reservoir (Fig. 2b,c). Its emission wavelength is temperature-tuned relative to the wavelength of the response laser, allowing us to control the reservoir properties[61]. In this implementation, the reservoir's delay loop is much longer than required due to the off-the-shelf fibre components used. Given the node separation of $\theta$=50ps, within one feedback loop we define in total *N*=1320 virtual nodes. We use only the first *k*=66 virtual nodes of the delay loop and we assign one sample from each bit's analogue pattern to one virtual node (*j=k*). In this way, the reservoir's delay line is only partially filled with a sequential input signal (Fig. 2d). The responses of the virtual nodes in the unused part of the reservoir are omitted from training, while the next bit pattern follows after one time delay ($\tau$) (Fig. 2e). The extended fibre length of the feedback loop has been assessed and does not affect the classification performance. In the presented scheme of sequential input feeding of the reservoir, 1-bit pattern from the transmission signal is assigned to one time delay of the reservoir (Methods). This means that 1-bit of information is time-stretched to fit into the k nodes of the reservoir's within one time delay. This offline time-stretch defines the so far remaining "speed penalty" of our processing method (Methods). The response output of the reservoir is recorded using a 80GSa/s real-time oscilloscope and is used to train and test the linear classifier (Methods). A weighted summation of reservoir responses provides the estimated bit sequence of the initial data. In the current investigation the classifier is calculated offline. However, the linear regression approach was selected since it can be implemented with hardware approaches that have been reported lately, based on all-optical implementations of temporal integrators[62,63] and fast analogue summation techniques[25].

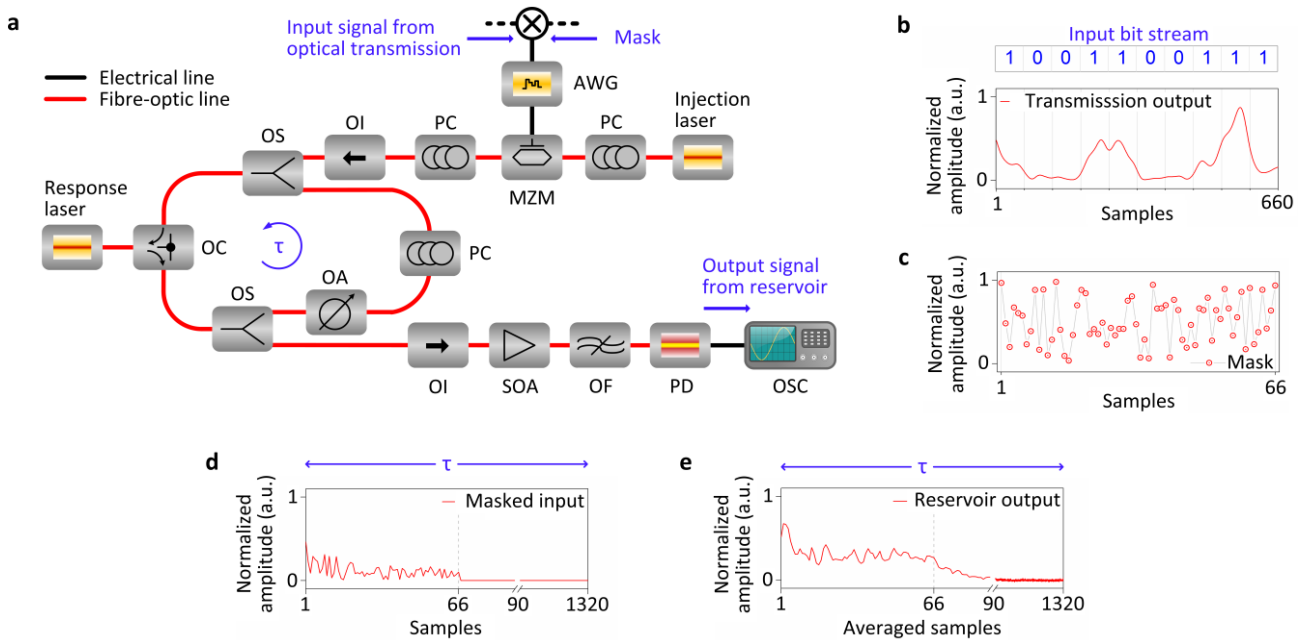

**Figure 2 | Experimental configuration and operation of the photonic reservoir. (a)** Scheme of the experimental setup. The response laser receives delayed optical feedback from a fibre loop realised using an optical circulator (OC). The feedback polarization is parallel to the response laser's emission and its strength is controlled by an optical attenuator (OA). The simulated signal after fibre-optic transmission and photodetection is randomly masked offline, uploaded into a 20GSa/s arbitrary waveform generator (AWG) with 10 GHz analogue bandwidth and applied on the injection laser's optical carrier via a 20GHz Mach-Zehnder modulator (MZM) that is operated in the linear regime. The optical output of the reservoir is amplified by a semiconductor optical amplifier (SOA), filtered (OF), photodetected (PD) and recorded by a 80GSa/s 16GHz analogue bandwidth real-time oscilloscope (OSC). PC: polarization controller. OI: Optical isolator. OS: Optical splitter. **(b)** Example of input bit stream with 10 bits. One bit per time delay is encoded and is

represented here by *j*=66 samples. This representation generates a speed penalty in processing. **(c)** Mask profile. The *j* samples are masked sequentially with *k*=66 (*j*=*k*) values in the interval [0,1] (Methods). **(d)** The long reservoir time-delay ($\tau$=66ns), along with the fast sampling of the AWG (20GSa/s), allows for up to *N*=1320 virtual nodes. In our task, we use only *k*=*N*/20 of them here: the masked input per bit is assigned to the first 66 virtual nodes, while the subsequent bit follows after *N-k*=1254 zero samples. **(e)** The acquisition of the output at 80GSa/s allows us to average four detected samples per virtual node and increase the signal-to-noise ratio. The reservoir's response per bit contained in the first 66 nodes (66 samples after averaging) is considered for the training the classifier.

**Results and discussion**

In bit streams without temporal cross-talk, a classifier would be trained on the currently evaluated bit only, by solely considering its timeframe. Nevertheless, the presence of deterministic fibre transmission impairments results in patterns that contain information from neighbouring bits also. Contemporary DSP methods commonly use this information by considering one sample per bit in order to improve detection capabilities. Here we follow the same approach, but using multiple samples per bit – the patterns of the neighbouring bit timeframes – in order to improve classification performance (Methods). The optimal number of timeframes to be considered for training depends on the extent of chromatic dispersion and the Kerr nonlinearity. In the notation we adopt, training the classifier with *n*-bits means that we consider a reservoir's response of duration $n \cdot \tau$ that includes those bit timeframes as defined in Fig. 3a. The number of transient responses that participate in the training is then $n \cdot k$.

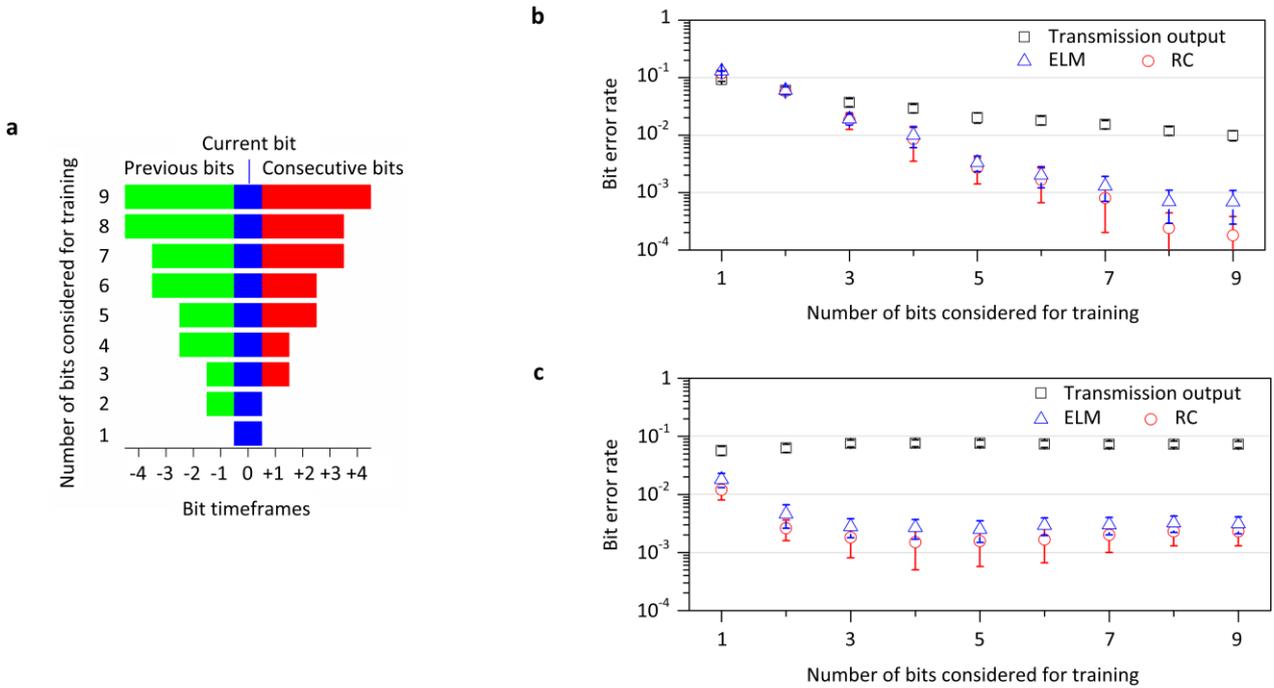

**Figure 3 | Signal recovery by training an experimentally implemented photonic reservoir. (a)** Schematic that associates the number of bits considered for training with the corresponding bit timeframes of the reservoir response that is taken into account. In order to take into account the consecutive bits in the computation, a time latency of the processing applies, equal to the total duration of the

consecutive bits. In this case, the latency is 4-bit timeframes (4τ). **(b)** BER performance of the recovered test bit stream versus the number of bits considered for training, when handling degraded signals from the short-reach transmission system (Supplementary Fig. 1) and considering different processing methodologies: training on the transmission output as received from the communication channel (black rectangles), training on the output response of the photonic reservoir with minimized optical feedback (ELM approach - blue triangles) and training on the output response of the photonic reservoir with optimized optical feedback conditions (RC approach – red dots). **(c)** The same as **(b)**, but for the long-haul transmission system (Supplementary Fig. 2). BER statistics emerge from 5 independent repetitions of the experiment that use different bit streams as input.

*Short-reach transmission*

When considering the short-reach transmission scenario, an extended distance of $z_1$=45km leads to a BER value above 0.1 (incoherent detection in absence of any DSP). Training the linear classifier directly on the output signal from transmission (benchmark test of Fig. 1d) within 1-bit timeframe, the BER measured for the test sequence is 0.1 (Fig. 3b, black rectangles). Even though the pattern within one bit duration provides more information than a single sample in the same duration, BER is not reduced significantly. When considering for training an extended sequence of 9-bits, the BER is reduced to $10^{-2}$, driven by the effect of the pattern recognition that uses information from neighbouring bits. Equivalent results are obtained for the benchmark classification test of Fig. 1c. In these tests, the number of samples per bit is equal to the number of nodes that will contribute to the reservoir computation ($j=k=66$). Thus, for 9-bit training, 594 reservoir outputs contribute to the linear regression model. When we incorporate the reservoir in the system and optimize its performance with respect to feedback strength and laser frequency detuning (Supplementary Fig. 5), we obtain significantly improved BER values as low as $1.8 \cdot 10^{-4}$ (Fig. 3b, red dots). Even when minimizing the feedback in the reservoir loop and operating the system as an ELM, we still obtain an improvement compared to the benchmark tests, with a BER value as low as $7 \cdot 10^{-4}$ (Fig. 3b, blue triangles). These findings are very encouraging and indicate two contributing mechanisms of the photonic reservoir to the improvement of the binary classification performance. The first one is attributed to the nonlinear transformation of the injected signal into the response laser (ELM operation) and the second one is attributed to the inherent fading memory[61] of the reservoir (RC operation). Thus, when comparing a temporal pattern of 9-bit duration (4 previous, the current and 4 subsequent bits) we obtain significant improvement compared to the benchmark tests of Fig. 1c and 1d. Nevertheless, the optimal number of the neighbouring bits we may consider is not fixed; it is related to the extent of transmission impairments and the transmission length, thus it may be selected accordingly.

*Long-haul transmission*

As a second task, we test our scheme considering a long-haul communication link with $z_2$=4000km, following the same methodology. Training directly on the transmission output signal we obtain for 1-bit training a BER value as high as 0.056 (Fig. 3c, black rectangles). By considering the photonic reservoir, but without masking the input (benchmark test of Fig. 1c), we obtain the same performance. However, after masking the input signal and optimizing the reservoir's operating conditions, the BER value we measure is significantly reduced to $1.7 \cdot 10^{-3}$. Here, it is obtained considering only 4-bit or 5-bit timeframes for training (Fig. 3c, red dots and Supplementary Fig. 6). In this transmission scenario, the number of neighbouring bits that affect the current bit profile is smaller, since chromatic dispersion compensation is applied to this system. Also in this type of links, the presence of stochastic optical amplification noise (here we considered an optical amplification noise figure of 5dB) is an additional limiting factor for data recovery. Reservoir computing cannot compensate for such stochastic processes. Yet again, the RC approach with optimized feedback conditions yields significant nonlinearity mitigation, with slightly better results than the ELM approach (Fig. 3c, blue triangles).

*Extension of communication range*

The found improvements of the BER level of the detected signals can be directly translated into an excess in usable transmission distance. We focus on BER values around $10^{-3}$; at this decoding BER threshold, a hard-decision forward error correction (FEC) method can provide an error-free data recovery. FEC codes impose an overhead to the initial data sequence, depending on the data bit-rate (typically 12% for $R_2$ and 7% for $R_1$)[64]. In this work, we consider the initial data rates. Evaluating the short-reach transmission scheme, as an example, we obtained a BER improvement of almost two orders of magnitude with respect to the benchmark tests. The resulting gain in transmission distance using the RC post-processing is 75.9% when compared to classifying the transmission output and 200% when compared to the direct detection performance without any processing (Fig. 4). These are remarkable extensions, illustrating the power and potential of the presented approach, extracting the bits via a pattern recognition method based on photonic reservoir computing.

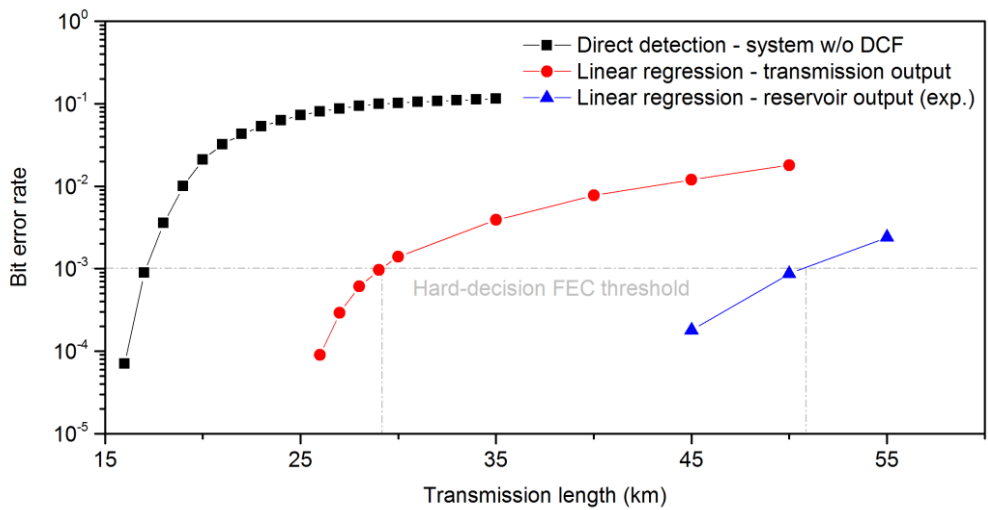

**Figure 4 | Bit error rate dependence on fibre transmission distance in the short-reach communication system.** When considering direct detection of the transmitted signal (black rectangles), the BER limit for hard-decision FEC is met for a fiber

transmission length equal to 17km. When applying the linear regression algorithm with 9-bits of training to the transmitted signal (red dots), the transmission length is extended to 29km. Finally, when applying the photonic reservoir and the same algorithmic training (blue triangles), the transmission length is extended further to 51km. The latter performance is obtained from the experimental implementation of the photonic reservoir. Therefore it includes all noise sources and instabilities of the system to environmental perturbations.

*Towards faster reservoirs*

Reservoirs with much smaller delays than the one demonstrated here can be designed by exploiting photonic integrated circuits, without losing classification performance. These designs can take a priori into consideration the appropriate number of virtual nodes and transient states that will be needed for optimized training. In order to evaluate the limitations imposed by the reservoir size, we numerically simulate the behaviour of short reservoirs with the same virtual node spacing ($\theta$=50ps) and various feedback time-delays ($\tau$). We simulate a system that is analogous to the experimental topology of Fig. 2, based on a rate equation model for the response laser dynamics (Methods). Phase dependencies within the short feedback loops are also taken into account. As input signal to the reservoir, we consider the output signal from the short-reach transmission scenario we used before, but for a slightly increased transmission distance ($z_1'$=50km). First, we shorten the reservoir delay loop to $\tau$=1.6ns ($k$=32) and use an equal number of samples to describe each bit pattern ($j$=32). We obtain data recovery without any errors in the test bit sequence for several operating conditions of the reservoir (Fig. 5). This performance is preserved even when the input signal sampling is reduced to only $j$=4 samples per bit duration. In all cases that $j<k$, each sample contributes as input to more than one virtual node (Methods). The sampling rate of $j$=4 for 25Gb/s pulses is possible to obtain with the current state-of-the-art detection equipment, eliminating the need for oversampling. Further sampling reduction (eg. $j$=2) drastically degrades the classification performance. This indicates that pattern profiles within each bit duration include critical information for the classification tasks. The speed penalty of the current approach, for the case of $\tau$=1.6ns reservoir, is $\tau/R_1^{-1}$ =1.6ns/40ps=40 (Methods). This speed penalty can be further reduced when considering even shorter delays in the proposed time-multiplexed approach. Nevertheless, numerical simulations show that this comes at the expense of BER improvement (Fig. 5). A combination of our presented time-multiplexing approach with spectral or spatial encoding could overcome this restriction.

For the case of $\tau$=1.6ns, we measure the obtained BER as a function of the received optical power after a fibre transmission length of $z_1'$=50km. When considering a direct detection system without including any dispersion compensation, the BER threshold of $10^{-3}$ – where FEC techniques can apply – can never be reached. Even when considering a linear classifier with 9-bit training on the received signal, the previous threshold is still not reached (Fig 6, blue triangles). On the contrary, the BER threshold is reached when considering the same classifier with 9-bit training on the photonic reservoir output and at a received power as low as -17.5dBm (Fig 6, red dots). As a reference for comparison to this performance, we show in Fig. 6 (black rectangles) the performance of an optimized transmission system that includes also physical dispersion compensation (i.e. DCF), for the same transmission distance $z_1'$. Even though we determine a power penalty of 5.8dB when considering the RC-based detection, we show that this method has a remarkable potential to mitigate both, linear and nonlinear transmission phenomena. The performance of the reservoir of this case ($\tau$=1.6ns, $z_1'$=50km) and for reservoir sampling $(j,k)=[4,32]$ has been also evaluated as a function of the optical SNR of the received signal after transmission (Supplementary Fig. 7).

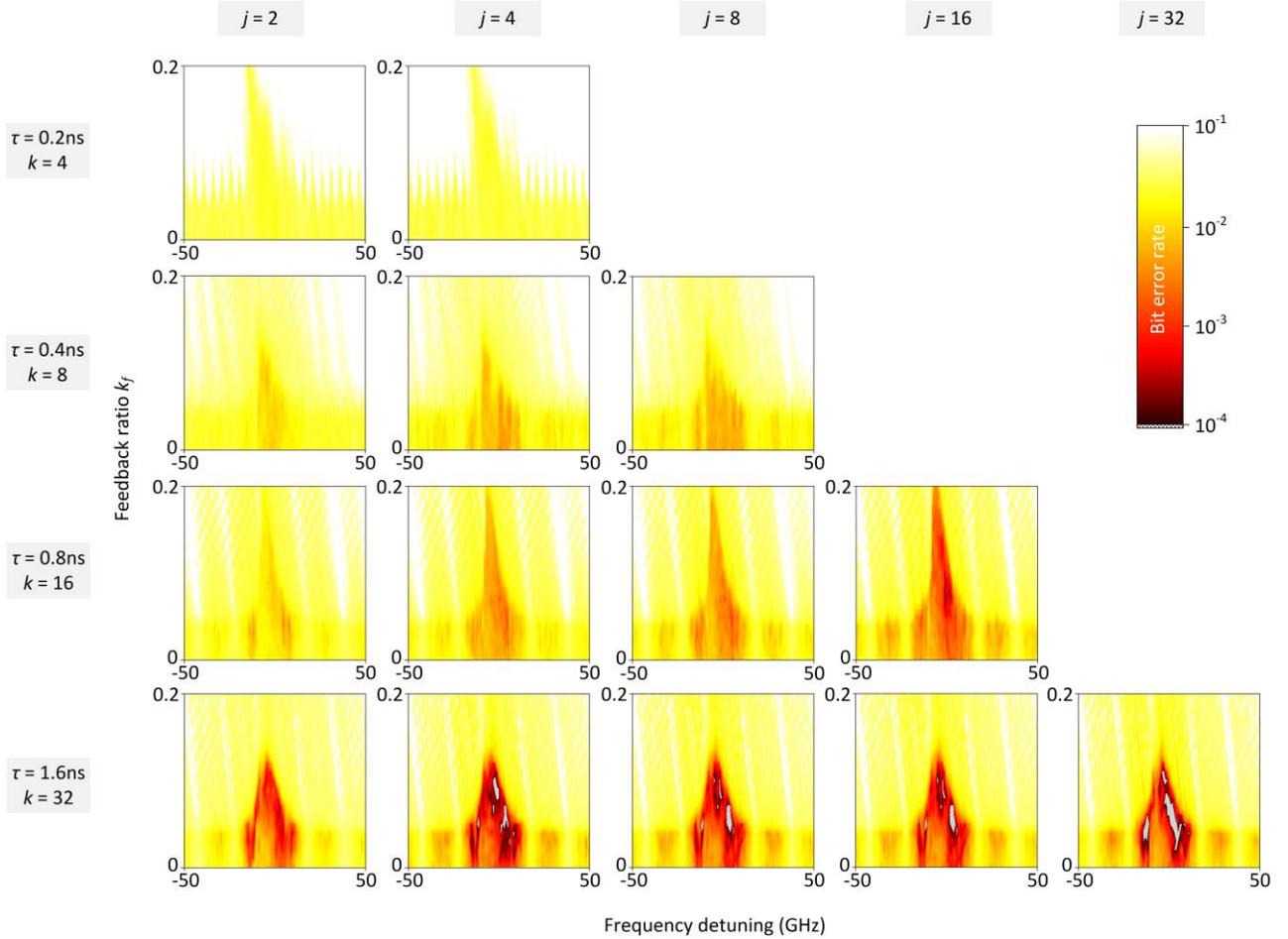

**Figure 5 | Simulated RC performance of a reservoir with short time delays.** The number of samples used per bit representation (*j*) is disengaged from the number of virtual nodes (*k*) per time delay $\tau$ that are used for training (Methods). The input sequence tested here is for the case of the short-reach transmission system and for $z_1'$=50km (Supplementary Fig. 1,3). Training is performed on a 9-bit reservoir response, as presented in Fig. 3a. Here we map the data recovery performance in BER terms of the test bit stream, by changing two parameters of the reservoir operation: the frequency detuning $\Delta f$ between the injection and the reservoir laser and the feedback ratio $k_f$ (Methods). There is a significant dependence of the BER performance on the transient states used for training and the samples used for the bit representation. For $k$<32, RC processing fails to provide an error-free data recovery, even when we incorporate more samples of the input stream. Still, RC provides a significant BER improvement as long as we are using a considerable number of transient states in the training. For comparison, when we apply 9-bit training without the use of the reservoir, BER of the recovered bit stream is higher than 0.02 for all (*j,k*) representations.

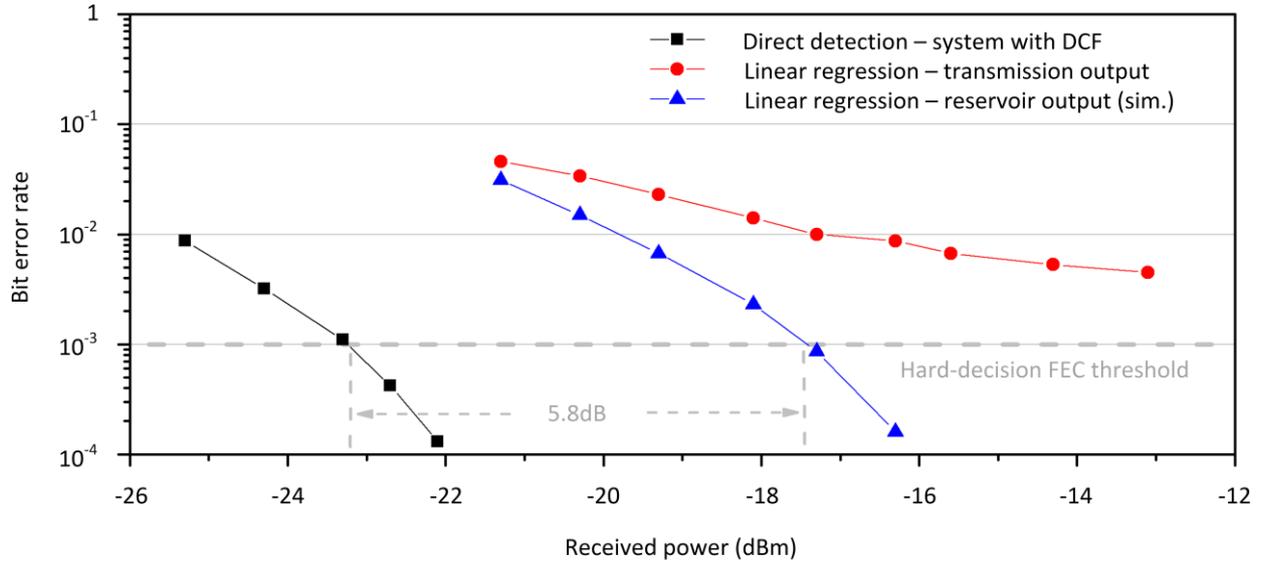

**Figure 6 | Bit error rate dependence on received optical power in a short-reach communication system with $z_1'$=50km.** The dependence on the received power is obtained by attenuating the optical signal before photodetection (Supplementary Fig. 1). We compare the BER performance, when considering a linear classifier with 9-bits of training (red dots) of the transmitted signal and the same classifier with 9-bits of training (blue triangles) of the photonic reservoir output. In both cases the launched optical power for transmission is 10mW. An analogous transmission system with direct detection that includes also physical dispersion compensating fibre (DCF) is shown as a reference (black rectangles). The optimal performance of the last system has been obtained for a launched optical power of 1mW.

*Conclusions*

Following a pattern recognition approach, our photonic RC-based hardware platform can efficiently process fibre transmission signals that suffer from deterministic transmission impairments. The adopted processing concept is generic and might be applied to various contextual pattern recognition tasks. In the context of communication systems, it can be extended for transmission systems with advanced modulation formats at even higher bit-rates. The concept can also be applied to transmission systems with quadrature modulation formats or coherent receivers. Information that may be encoded in the phase space can be easily converted into an amplitude signal and fed into the photonic reservoir as a microwave modulating signal, as presented in Fig. 2. This approach has proven to be efficient since the reservoir's states are not disturbed by the phase properties of the transmission signal. Moreover, the polarization state of the received signal after transmission does not interfere with the active polarization state of the reservoir. A future challenge of the proposed scheme is to extend it to a wavelength division multiplexed (WDM) transmission environment, with the presence of nonlinearities that originate from neighbouring channels. The high launched optical power conditions we considered in this work will induce four-wave mixing effects that might degrade the detection performance. The training might need to be performed also on the patterns of neighbouring channels that affect a specific channel's bit sequence. In a scenario like this, the required training data sets are expected to be significantly larger in order to achieve efficient training. It is an open question to what extent the photonic reservoir's nonlinear transformation and fading memory will offer improved detection capabilities. Finally, the presented time-multiplexing approach that feeds the input signal into the reservoir represents only one possible coding method. Complementary methods – based on spectral or spatial multiplexing[65] – for high-dimensional mapping of the input to the reservoir states are envisaged to minimize speed penalty and eventually lead to real-time binary classification.

**Methods**

**Numerical simulation of fibre-optic transmission systems**

We model numerically the fibre transmission using a coupled nonlinear Schrödinger equation (CNLSE) propagation model[28,66], in the presence of two orthogonal polarization modes, including also fibre attenuation, chromatic dispersion, Kerr nonlinearity, inter-channel nonlinear effects (cross-phase modulation and four-wave mixing) and stimulated Raman scattering. However, some of the above phenomena are not activated in our investigations due to the selected properties of the transmission systems. For example, we consider only single channel transmission; thus inter-channel nonlinear effects are not present. Moreover, the output optical signal from transmission is photodetected and the electrically converted signal is used as an input to the photonic reservoir. In this case, the reservoir system becomes robust to the polarization state of the received optical signal. Hence, for sake of simplicity we write the nonlinear Schrödinger equation in a form that describes only the phenomena that affect critically the transmitted signal. The slowly varying optical field $E_{tr}(t,z)$ that travels along the SSMF is given then by:

$$i\frac{\partial E_{tr}}{\partial z} + i\frac{a_{loss}}{2}E_{tr} - \frac{\beta_2}{2}\frac{\partial^2 E_{tr}}{\partial t^2} + \gamma|E_{tr}|^2 E_{tr} = 0 \qquad (1)$$

where $z$ is the distance in km, $t$ is the relative time in the frame that moves with the envelope velocity, $a_{loss}$ is the fibre transmission loss coefficient, $\beta_2$ is the chromatic dispersion coefficient and $\gamma$ is the instantaneous Kerr nonlinearity coefficient. We consider a DFB single mode laser that emits 10mW of optical power at 1550nm, with a linewidth of 0.1MHz. This launched power level induces Kerr nonlinearity in the transmission line. We consider a typical parameter set for the SSMF transmission: $a_{loss}$=0.2db/km, $\beta_2$=21.7ps$^2$·km$^{-1}$ and $\gamma$=3·10$^{-29}$km$^2$·mW$^{-1}$. In the long-haul transmission modules, the parameter set for the DCF is: $a_{DCF}$=0.6db/km, $\beta_{2,DCF}$=-128ps$^2$·km$^{-1}$ and $\gamma_{DCF}$=2.6·10$^{-29}$km$^2$·mW$^{-1}$. The amplification unit provides a gain of $G$=30.2dB, with a noise figure $NF$=5dB, while the optical filter has a Gaussian profile with 3-dB optical bandwidth equal to 4 times the signal bit-rate. The received optical signals are photodetected with a typical PIN photoreceiver, with responsivity of 0.9A/W. Thermal noise and dark current noise are included in the photodeteciton stage. Finally, the signals are electrically filtered with a low-pass, 4$^{th}$ order Bessel filter, at a cut-off frequency of 0.8 the data bit-rate. The numerically generated signals are used to feed our experimentally-built reservoir, as well as the numerical investigations of the RC operation.

**Input signal masking**

Each bit is represented by an analogue pattern of *j* samples. If *j=k*, each sample $a_j^i$ of the $i^{th}$ bit is masked by a random value $m_k \in [0,1]$ and the dimensionality of the corresponding state space is preserved. This condition applies for our experimental implementation. Conventionally, the classification performance in machine learning techniques is achieved by nonlinearly mapping the input onto a higher dimensional state space. The higher dimensionality results in a higher chance of linear separability of the pattern classes that need to be distinguished. Here, we demonstrate that for this mechanism to work it is sufficient to maintain the state space dimensionality defined by the number of samples. Nevertheless, due to the masking, we create a higher dimensional representation of the pattern in the state space. If the analogue input is sampled with fewer samples than the reservoir's virtual nodes (*j<k*) used for classification, each sample $a_j^i$ is masked with two or more mask values of the *m* vector. In this way the dimensionality of the state space is increased. The condition *mod(k,j)=0* should be preserved so that all samples are equivalently masked. When considering short reservoirs with a small number of virtual nodes, the random mask vector *m* consists only of a few values. Thus, the dependence of the system performance on the chosen mask becomes significant. In the numerical results shown in Fig. 4, the selection of the mask is critical – especially when *k≤16*. For this reason, 10 different random masks have been considered for evaluating the BER performance for each investigated *(j,k)* case. The maps presented in Fig. 4 refer to those masks that lead to the lowest BER values. In contrast, in all the experimentally investigated scenarios, the dependence of the BER performance on different tried random mask sequences was insignificant. When k value is small, i.e. a mask of just a few values vector, we work at a lower dimensional state space, and in this case the condition of selecting a random mask is lost. There are insights of selecting a good mask and there also optimization procedures have been proposed[67]. However, here we chose a random mask that consists of random values for reasons of simplicity.

**Experimental reservoir configuration**

All devices used in the experiment are commercial devices. The response laser is a discrete-mode quantum-well SL from Eblana Photonics emitting at 1542 nm, with a longitudinal mode separation of 145 GHz, side mode suppression ratio of 40 dB and threshold current of $I_{th}$ = 11.2mA. The response laser is biased at $I_{bias}$ = 11.1mA, corresponding to 1% below solitary threshold. For maximum feedback conditions of the reservoir (zero attenuation at the OA, within the feedback loop), the threshold current is reduced to $I_{th,fb}$ = 10.5mA. The injection laser is a DFB laser from Toptica emitting also at 1542 nm, with >30dB side mode suppression ratio and high power emission (up to 40mW). Temperature and bias current were stabilized with 0.01 K and 0.01 mA accuracy, respectively. The masked signal from transmission is uploaded on the emitted carrier of the injection laser through a 20 GHz Eospace Mach-Zehnder amplitude modulator (MZM). The output optical signal from the reservoir is amplified by a Covega semiconductor optical amplifier (SOA), filtered by a Santec OTF350 tunable optical filter and detected by a Miteq SCMR-100K20G avalanche photodiode with 20GHz bandwidth and 2kOhm transimpedance gain. Signal monitoring is performed by a Lecroy Wavemaster 816Zi Oscilloscope and an Aragon Photonics High-resolution optical spectrum analyser with 10 MHz resolution. Optical isolators with 50 dB isolation are included in the optical path to suppress unwanted reflections. The above information can be included in the manuscript.

**Training and testing**

We generate independent binary streams of 40960 bits, generated with random seeds. From these sequences, we use one stream for training and cross-validation evaluation and one stream as the testing set. From the first sequence, 75% of the bits for training and 25% are used as the cross-validation set. We use a ridge regression algorithm with 10 repetitions of Monte-Carlo cross-validation[68] and a ridge parameter of 0.01. The test set is finally used to evaluate the BER performance recovery of an independent data set, that the system was not trained on that. In this way we avoid possible pitfalls in biased training[69]. The reservoir's node responses $r_k^i$ that correspond to the $i^{th}$ bit duration are used to train the linear classifier. In this case we consider only 1-bit timeframe duration for training (the one of the currently predicted $i^{th}$ bit). However, the predicted bit stream has the generic form of the weighted sums of these responses: $\tilde{b}^i = \sum_{k,n} w_k \cdot r_k^n$, where $n$ indicates an extended number of neighbouring bit responses. Due to fibre nonlinearities, signal properties at neighbouring bit timeframes can include critical information that can be exploited by the classifier. $n=[i - m_p, i + m_c]$, with $m_p \in \mathbb{N}$ being the number of previous and $m_c \in \mathbb{N}$ being the number of consecutive bits, all used for the $i^{th}$ bit classification. The $i^{th}$ bit prediction is made with a latency of $m_c$ bits. In order to have access to the previous and consecutive bit timeframes, the bit stream considered is slightly shorter in length (40960-$m_p$-$m_c$). The optimum weights are found by minimizing the difference between $\tilde{b}^i$ and $b^i$ for all of the bits that are used in the training set. The optimal $w_k$ values are determined by using techniques for extracting eigenvalues from singular matrices such as the linear Moore–Penrose pseudo-inverse operator (denoted by †). If the target matrix $\tilde{B}$ contains the targets $\tilde{b}^i$ for all of the reservoir's loops $\tau$ that are used for the training, and the response matrix $S$ contains all reservoir's responses for the same loops, then the matrix $W = \tilde{B} \cdot S^\dagger$ contains the optimal weights. Training is performed offline on a computer and takes no longer than several seconds.

**BER calculation**

The BER is calculated from the predicted bit stream $\tilde{b}^i$, following the same procedure as for conventional communication systems. A decision threshold is set in a binary comparator aiming on BER minimization. Since 25% of the bit stream (10240 bits) is considered for cross-validation, we can measure a minimum BER value of ~9.8·10$^{-5}$ for the training performance. The independent data streams that are used as test sets are evaluated using all 40960 bits, allowing thus a minimum BER value of ~2.4·10$^{-5}$. Depending on the values of $m_p$ and $m_c$, the total length of the test bit sequence might be shorter. However, this does not affect the minimum BER we can obtain from our measurements. Our interest focuses at a BER levels around $10^{-3}$ – one order of magnitude higher than the minimum value we can measure – where hard-decision FEC methods operate and offer error-free operation.

**Numerical model of the photonic reservoir**

We implemented a reservoir nonlinearity that follows the Lang-Kobayashi rate equation model of a SL with time-delayed feedback[70], with an additional optical injection term with frequency detuning. An analytical description of the implemented model under optical feedback and optical injection of frequency detuned signals can be found in [71]. The modelled rate equations for the response SL operation are:

$$\frac{dE_r(t)}{dt} = \frac{1}{2}(1+ja)[G_r(t) - t_{ph}^{-1}] \cdot E_r(t) + \frac{k_f}{t_{in}} \cdot E_r(t-\tau)e^{j\omega_0\tau} + \frac{k_{inj}}{t_{in}} \cdot E_{inj}(t)e^{-j\Delta\omega t} + \sqrt{D}\xi(t) \quad (2)$$

$$\frac{dN_r(t)}{dt} = \frac{I}{e} - \frac{N_r(t)}{t_s} - G_r(t) \cdot |E_r(t)|^2 \quad (3)$$

$$G_r(t) = g_n \cdot [1 + s|E_r(t)|^2]^{-1} \cdot [N_r(t) - N_0] \quad (4)$$

The angular frequency detuning $\Delta\omega = 2\pi \cdot \Delta f$ is defined between the emission frequencies of the injection and the response laser. A Gaussian white noise $\xi(t)$ is included for the electrical field with amplitude $D=30\text{ns}^{-1}$. The bias current for all lasers is set to $I=15.3 \cdot 10^{-3}$A (just below threshold current of solitary operation $I_{th}=15.37 \cdot 10^{-3}$A), while the remaining set of parameters is: $a=3$, $s=5 \cdot 10^{-7}$, $N_0=1.5 \cdot 10^8$, $g_n=1.2 \cdot 10^{-5}\text{ns}^{-1}$, $t_s=2$ns, $t_{in}=10^{-2}$ns, $t_{ph}=2 \cdot 10^{-3}$ns, $e=1.602 \cdot 10^{-10}$A·ns. These parameters do not simulate exactly the SLs that were used in the experiments; however they describe the general performance of the system. The response SL's angular frequency is $\omega_0=2\pi c/\lambda_0$ where $\lambda_0=1550$nm and $c$ is the speed of light. We define an injection field with the parameters: $k_{inj}=0.15$ and $E_{inj,0}=100$, so that after conversion this corresponds to an injection optical power of 0.6mW. This coincides approximately with the optical power that we used in our experiment. When considering the modulation properties of the injection signal due to the masked input, the injected electrical field is of the form:

$$E_{inj}(t) = E_{inj,0} \cdot [b_{bias} + m(t) \cdot a(t)] \quad (5)$$

$m(t) \in [0,1]$ is the mask sequence, $a(t)$ is the photodetected signal after transmission (normalized also in the range [0,1]) and $b_{bias}$ is a bias term that ranges from 0 to 1 and controls the average optical power that is injected into the response laser. The feedback ratio $k_f$ in this model expresses the ratio of the response SL's emitted electrical field that is redirected back to the SL. As presented in Fig. 4, a feedback value of $k_f \sim 0.05$ results in an optimal operation of the reservoir. When converting this electrical field ratio into optical power attenuation terms with respect to the optical power emitted by the response SL, we get a feedback attenuation of ~26dB. This theoretically estimated value is consistent with the experimental optimal conditions we observed for the short-reach transmission system, but for the long reservoir delay. As shown in Supplementary Fig. 5, the optical attenuation for which we achieve the optimal BER performance (case of $9\tau$ training) is 12dB. The fibre loop itself has an initial loss of 7dB, due to optical splitters and fibre components included in it. An additional loss of ~6dB should be also considered from the dual-pass loss between SL facet and the fibre pigtail. In total, the experimental feedback attenuation is estimated to be ~25dB, very close to the numerically foreseen.

**Speed penalty**

In the simplified approach we adopt in this work, the input layer of the reservoir is implemented by feeding the input sequence using a time-multiplexing approach (sequential introduction of bits). Specifically, we assign the analogue pattern of one bit to be included within one feedback delay of the reservoir. This condition allows the inherent memory of the reservoir to introduce connectivity between subsequent bits (samples that are one $\tau$ apart in time), but at the same time induces a speed penalty. In principle, the bit durations ($R^{-1}$) we are handling in optical communication systems are much smaller than the reservoir's time-delay $\tau$. In this work we make a time-stretch of the bit time duration $R^{-1}$ to fit in one $\tau$ duration offline. The speed penalty of this RC post-processing step is defined as $\tau/R^{-1}$. In practice, several time-stretching methodologies[72] could apply for an actual implementation of this step.

**Acknowledgments**

We thank Claudio R. Mirasso, Miguel C. Soriano, Daniel Brunner, Moritz Pflüger and Silvia Ortín for helpful discussions. This work was supported by the Ministerio de Economía y Competitividad and FEDER via project IDEA (TEC2016-80063-C3), and by the European Union's Horizon 2020 research and innovation programme under the Marie Skłodowska-Curie contract 707068.

**Author contribution**

A.A. and I.F. developed the idea, A.A., J.B. and I.F. planned the experiments, J.B. and A.A. performed the experiments, A.A. did the data analysis and A.A. and I.F. wrote the manuscript.

**Competing financial interests**

The authors declare no competing interests.

**Materials & Correspondence**

The data that support all findings in this study are available from the corresponding author on reasonable request. Other requests for materials should be addressed to A.A. and I.F.

**Supplementary Figures**

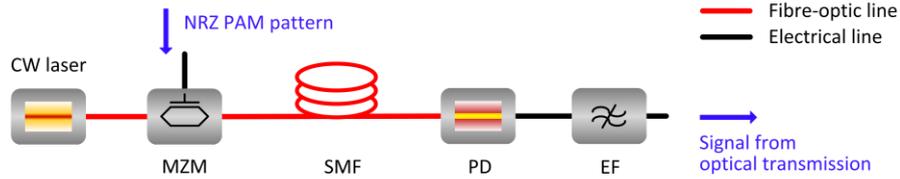

**Supplementary Fig. 1. Short-reach transmission system.** A continuous-wave SL, with relative intensity noise (RIN) of -145dB/Hz, emits 10mW of optical power and is externally modulated by a Mach-Zehnder modulator with 30dB dynamical range. The modulation signal is a random PAM, NRZ pattern at $R_1$=25Gb/s. A typical ITU-T G.652 SSMF is considered for transmission ($z_1$), while the optical signal is detected by a PIN photodetector. A 4$^{th}$ order Butterworth electrical filter with 20GHz bandwidth (0.8·$R_1$) is used to filter out high frequency signal components.

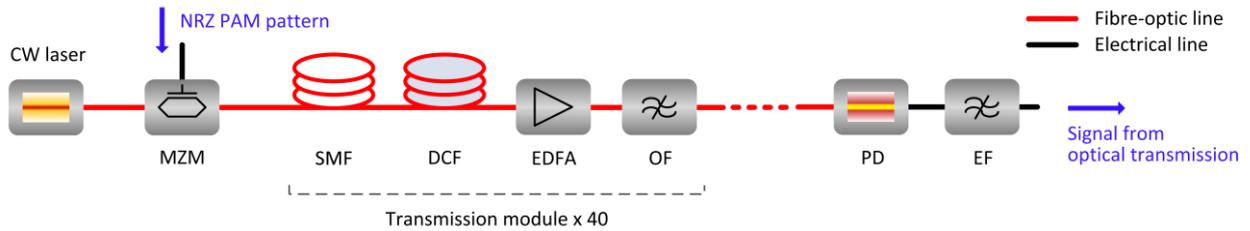

**Supplementary Fig. 2. Long-haul transmission system.** For this system, the same SL emitter, MZM and modulation pattern are used, but the bit rate of the data stream is now $R_2$=10Gb/s. The transmission line consists of 40 serially-repeated identical transmission modules. Each module consists of 100km-long SSMF, an appropriate length of dispersion compensation fibre that is matched to cancel completely the chromatic dispersion effects, an erbium doped fibre amplifier (EDFA) that compensates the power loss of the transmission path and an optical filter with $4·R_2$ optical bandwidth. In our considerations, the total length of SSMF transmission is $z_2$=4000km, without including the length of DCF. The detection stage is the same as in the configuration of the short-reach transmission system, while the electrical filter bandwidth is now reduced to 8GHz (0.8·$R_2$).

*arXiv:1710.01107v4*

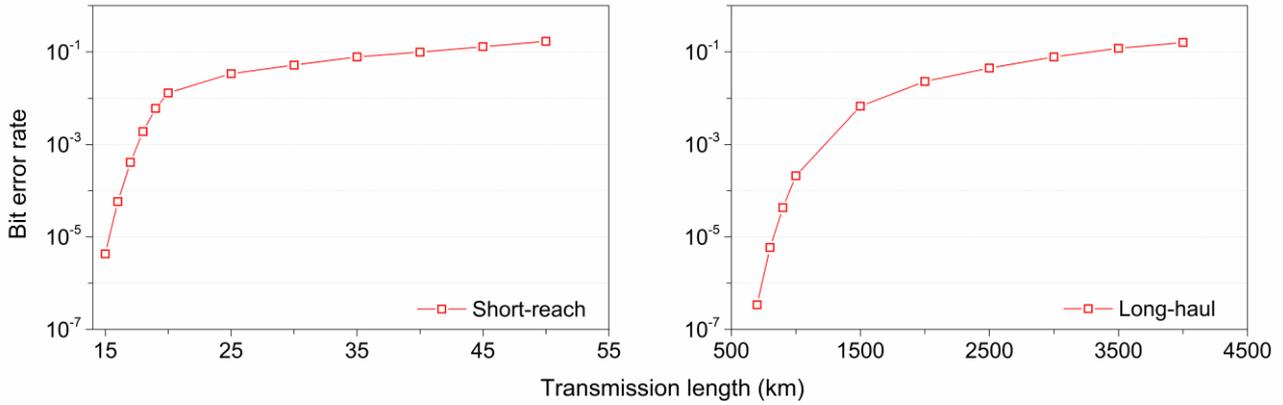

**Supplementary Fig. 3. BER versus transmission length for shot-reach and long-haul transmission systems.** For the transmission configurations presented in Supplementary Figs. 1 and 2, we estimate numerically the BER performance by considering different transmission lengths in SSMF. For the case of long-haul transmission we consider different number of transmission modules, by preserving the same internal structure. The BER of the decoded signal is higher than 0.1, for $z_1$>41km in the case of short-reach transmission and for $z_2$>3300km in the case of long-haul transmission. This consideration does not include any signal equalization with post-processing techniques.

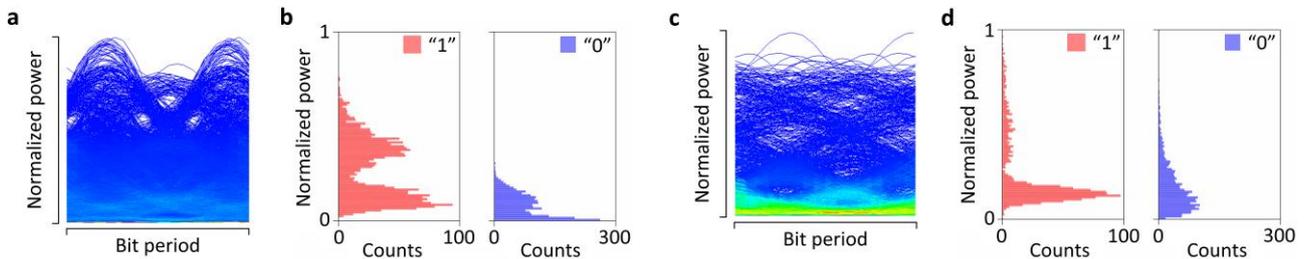

**Supplementary Fig. 4. Properties of the photodetected signals after transmission.** (**a**) Eye-diagram and (**b**) histogram of the detected power of the bit stream sampled at 0.6 of the bit period, for the short-reach transmission system at $z_1$ =45km distance. Chromatic dispersion and Kerr nonlinearity eliminate the potential for an efficient binary level separability. (**c**) Eye-diagram and (**d**) histogram of the detected power of the bit stream sampled at 0.6 of the bit period, for the long-haul transmission system at $z_2$ =4000km distance. Kerr nonlinearity and spontaneous emission noise from amplification units eliminate the potential for an efficient binary level separability.

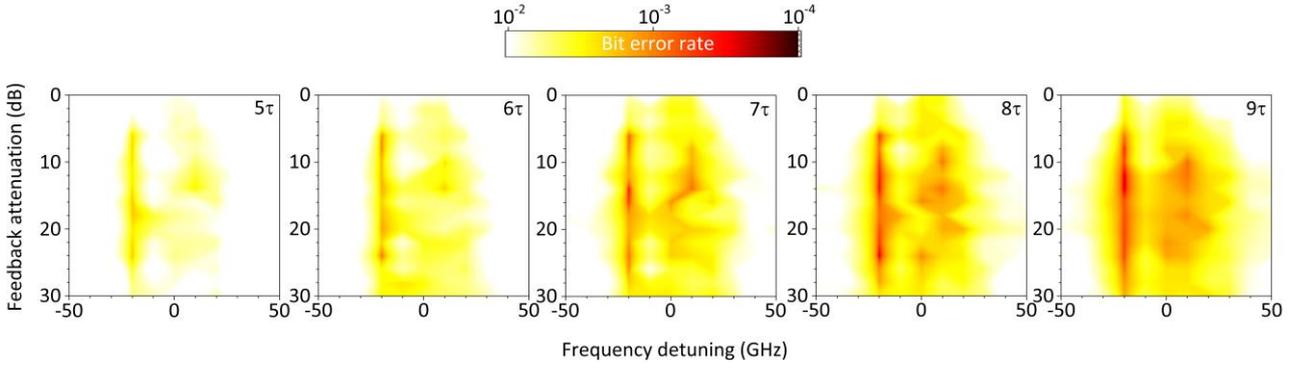

**Supplementary Fig. 5. Experimental BER mapping of the short-reach system after training on the reservoir output.** The mapping is performed versus frequency detuning *Δf* between the injection and the reservoir laser, while the feedback strength is controlled by attenuating the feedback that is available at the reservoir. Zero attenuation means that the feedback strength is defined by the setup structure and in our case this is roughly 20% (7dB less) of the emitted optical power of the reservoir SL. Moreover, the mapping is repeated for various numbers of bit timeframes taken into account for the linear classification (from $5\tau$ to $9\tau$). In the short-reach transmission case, bit information is shown to spread up to 4 neighbouring bits. Each map is the average of the BER values taken from 5 independent measurements. The transient states of the photonic reservoir depend directly on its operating conditions and the injected signal. In our classification task, their properties determine the BER performance of the recovered signal. Complete injection locking conditions that appear around *Δf=-10GHz* for a wide range of feedback values, result in poor BER improvement. Completely unlocked conditions, observed for frequency detuning values below *Δf =-30GHz* and above *Δf =30GHz*, also result in poor performance. However, for partial locking regimes and moderate feedback conditions the system exhibits significant BER improvement. This improvement is observed when considering at least a $5\tau$ timeframes response from the reservoir and is maximized for $9\tau$. The optimized BER is measured to be $1.8\cdot10^{-4}$ and is recorded for *Δf=-20GHz* and an excess feedback attenuation of 12dB. By minimizing the effect of feedback (30dB excess attenuation) we still get improved classification (BER=$7\cdot10^{-4}$) compared with the benchmark tests. In this case the reservoir is degenerated to an ELM that benefits from the nonlinear transformation of the injected input to the semiconductor laser.

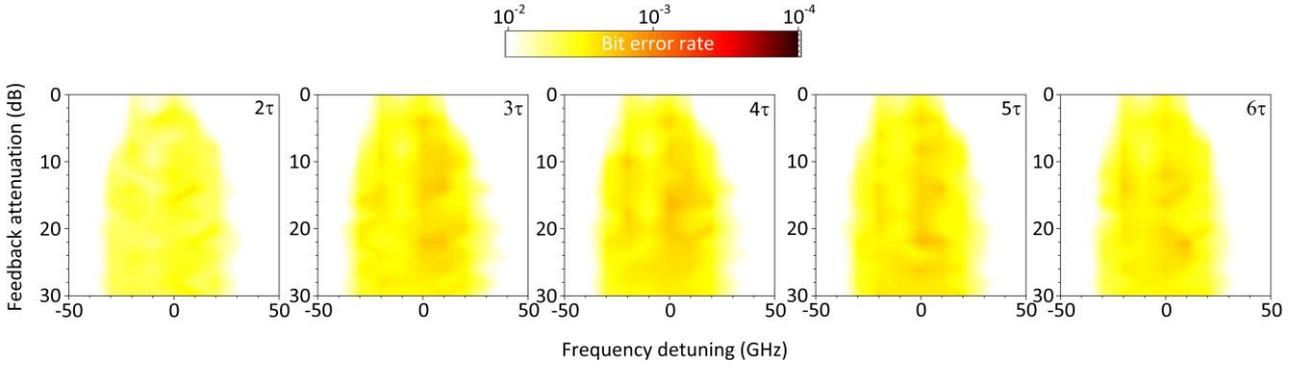

**Supplementary Fig. 6. Experimental BER mapping of the long-haul system after training on the reservoir output.** An analogous mapping with the one presented in Supplementary Fig. 4 is performed for the long-haul transmission case. In this case, the mapping is repeated for training with shorter bit timeframes (from $2\tau$ to $6\tau$), since bit information spreads only to just one or two neighbouring bits, due to the dispersion compensation of the signal along the transmission line. Thus, the consideration of more virtual node responses from extended bit time-frames does not provide any additional information for the classification. Consistent observations regarding the locking conditions are also made here. Again, complete injection locking conditions that appear around $\Delta f=-10GHz$ for a wide range of feedback values, result in poor BER improvement. We find the optimal operating condition at $\Delta f=0GHz$ and at an excess feedback attenuation of 16dB, when considering a $4\tau$ timeframes response from the reservoir; the measured BER is $1.7 \cdot 10^{-3}$.

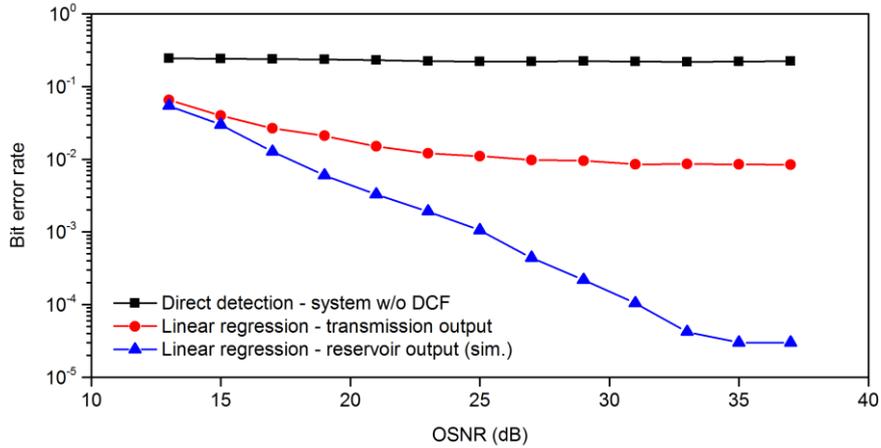

**Supplementary Fig. 7.** Bit error rate dependence on the optical SNR of the received signal in a short-reach communication system with $z_1'=50$km. The reservoir properties that have been considered are: $\tau=1.6$ns, $k=32$, $j=4$, $\Delta f=5$GHz and $k_f=0.05$. This corresponds to a BER value $<10^{-4}$, as shown in Fig. 5 of the manuscript. The dependence on the optical SNR is obtained by adding optical white noise before photodetection. We compare the BER performance, when considering a linear classifier with 9-bits of training (red dots) of the transmitted signal and the same classifier with 9-bits of training (blue triangles) of the photonic reservoir output. In both cases the launched optical power for transmission is 10mW. The analogous transmission system with direct detection without dispersion compensation is shown as a reference (black rectangles), it suffers from linear and nonlinear distortion and exhibits a BER>0.2 for the whole investigated range of optical SNR.